\documentclass[aoas,nameyear,dvips]{arximspdf}

\doi{10.1214/10-AOAS346}
\volume{4}
\issue{1}
\pubyear{2010}
\firstpage{1}
\lastpage{4}


\begin{document}
\begin{frontmatter}

\title{Introduction to papers on the modeling and analysis of network data}
\runtitle{Introduction to Section on Network Modeling}

\begin{aug}
\author[A]{\fnms{Stephen E.} \snm{Fienberg}\corref{}\ead[label=e1]{fienberg@stat.cmu.edu}}
\runauthor{S. E. Fienberg}
\affiliation{Carnegie Mellon University}
\address[A]{Department of Statistics and\\ \quad Machine Learning
Department\\ Carnegie Mellon University\\
Pittsburgh, Pennsylvania 15213\\USA\\ \printead{e1}} %adresu isvedimo
%komanda gale!
\end{aug}

% HISTORY:
\received{\smonth{3} \syear{2010}}

% ABSTRACT

% KEYWORDS

\end{frontmatter}

In today's world, networks seem to appear everywhere. There are social networks,
communication networks, financial transaction networks, gene
regulatory networks, disease transmission networks, ecological food
networks, mobile telephone and sensor networks and more. We, our
professional colleagues, our friends and family, and especially our
students, are often part of online networks such as \textit{Facebook},
\textit{LinkedIn} and now \textit{Google Buzz}.
Some network structures are static and others are dynamically evolving.
Networks are usually represented in terms of graphs with the nodes
representing entities, for example, people, and the edges representing
ties or relationships. Edges may be directed or undirected depending on
the application and substantive question of interest. In terms of
statistical science, a network model is one that accounts for the
structure of the network ties in terms of the probability that each
network tie exists, whether conditional on all other ties, or as
considered part of the distribution of the ensemble of ties.

Ideas and language from graph theory abound in the technical literature
on networks. A typical representation involves a network with $N$
nodes, having $\left({{N} \atop {2}}\right)$ unordered pairs of nodes, and hence $2\left({{N} \atop {2}}\right)$ possible directed edges. If the labels on edges reflect the
nodes they link, as $(i,j)$, $Y_{ij}$ represents the existence of an
edge from individual $i$ to $j$, and $\{\mathbf{Y}\} = \{ Y_{12},
Y_{13}, \ldots, Y_{(N-1)N} \}$ represents the ties in the graph. The
simplest network models assume the edges to be independent, while a
statistically more interesting class of models treats the dyadic
structures for pairs of nodes to be independent.

In an extensive review of the statistical literature on network
modeling, Goldenberg et al. (\citeyear{Gold2010}) note:%.~\cite{Gold2010} note:
\begin{quote}
Almost all of the ``statistically'' oriented literature on the
analysis of networks derives from a handful of seminal papers. In
social psychology and sociology there is the early work of Simmel
(\citeyear{Simm1950}) at the turn of the last century and Moreno (\citeyear{More1934}) in the 1930s,
as well as the empirical studies of Milgram (\citeyear{Milg1967}) and Travers and
Milgram (\citeyear{TravMilg1969}) in the
1960s; in mathematics/probability there is the {{Erd\"{o}s--R\'{e}nyi}}
work on random
graph models [Erd\"{o}s and R\'{e}nyi (\citeyear{ErdoReny1959}, \citeyear{ErdoReny1960}), and a closely
related \textit{Annals of Mathematical Statistics} paper by Gilbert
(\citeyear{Gilb1959})]. There are of course other
papers that dealt with these topics contemporaneously or even earlier.
But these are the ones that appear to have had lasting impact.
\end{quote}

Statistical work in the late 1970s and early 1980s emphasized models
that exploited dyadic independence, for example, in the work of Holland
and Leinhardt (\citeyear{HollLein1981}). More complex exponential random graph models
(ERGMs) then drew considerable attention; for example, see Frank and
Strauss (\citeyear{FranStra1986}). But the estimation of parameters for such models turns
out to have been more problematic than expected; for example, see the
discussion in Rinaldo, Fienberg and Zhou (\citeyear{RinaFienZhou2009}).

The network modeling literature has ``taken off'' in the past decade, in
part because of the interest in structures associated with the
internet, and there are contributors from many different disciplines,
including biology, computer science, statistical physics, sociology
and, of course, statistics. Kolacyzk (\citeyear{Kola2009}) provides a book length
treatment of a selection of approaches and Airoldi et al. (\citeyear{AiroBleiFienGoldXingZhen2007})
provides a compilation of relevant papers. In addition there is the
probabilistic literature that has derived from the Erd\"{o}s--R\'
{e}nyi--Gilbert formulations much of which is described in Chung and Lu
(\citeyear{ChunLu2006}) and Durrett (\citeyear{Durr2006}).

Methods for the analysis of network data now take at least as many
forms as the applications in which they arise. While the original
examples of networks analyzed in the literature were typically small
(e.g., $n=18$ nodes corresponding to monks in a monastery), the size of
networks analyzed with more modern methodology has grown exponentially.
Networks with 1000 nodes are common, for example, in the study of
protein--protein interaction, and online networks such as \textit{Facebook}
include hundreds of million nodes. An interesting statistical question
we can ask is whether there is a relevant asymptotics associated with
network models as we move into such high dimensions. A recent paper by
Bickel and Chen (\citeyear{BickChen2009}) %~\cite{BickChen2009}
opens the door to such important statistical issues by linking back to
ideas in the probabilistic network literature.

The response to our initial call for papers on the topic of network
modeling was so overwhelming that
we are dividing the special section into two parts, with the first
appearing in this issue of \textit{The Annals of Applied Statistics}
(Volume 4, No. 1), and the remainder in the next issue (Volume 4, No. 2).

In Part I of this special section, we include a diverse collection of
papers with applications spanning sampling of rare populations,
internet flows, gene networks, online e-loyalty networks,
document-as-nodes links induced from text, and more. The methodologies
begin with ERGMs but include sparse regression models and state space models.
\begin{itemize}
\item In \textit{Modeling Social Networks from Sampled Data}, Handcock and
Gile develop the conceptual and computational statistical framework for
likelihood inference for ERGMs based on sampled network information,
especially for data from adaptive network designs. They motivate and
illustrate these ideas by analyzing the effect of link-tracing sampling
designs on the collaborative working relations
between 36 partners in a New England law firm.

\item In \textit{Analysis of Dependence Among Size, Rate and Duration of
Internet Flows},
Park, Hern\~{a}ndez-Campos, Marron, Jeffay and Smith use Pearson's
correlation coefficient and extremal dependence analysis to study the
flows of packet traces from three internet networks.
The correlations between size and duration turn out to be much smaller
than one
might expect and can be strongly affected by applying
thresholds to size or duration. Using extremal dependence analysis,
they draw a similar
conclusion, that is, near independence for extremal values of
size and rate.

\item Peng, Zhu, Han, Noh, Pollack and Wang work with sparse
regression approaches in \textit{Regularized Multivariate Regression for
Identifying Master Predictors with Application to Integrative
Genomics Study of Breast Cancer}. They apply their methods to genome
wide RNA transcript levels and DNA copy numbers were measured for 172
tumor samples.

\item In \textit{Optimal Experiment Design in a Filtering Context with
Application to Sampled Network Data}, Singhal and Michailidis examine
the problem of optimal design in the context of filtering multiple
random walks on networks. They apply their methodology to tracking
network flow volumes using sampled data where the design variable
corresponds to controlling the sampling rate, and they relate their
approach to the steady state optimal design for state space models.

\item Political networks and gene regulatory networks are the primary
focus of application in \textit{Estimating Time-Varying Networks} by
Kolar, Song, Ahmed and Xing. They describe an approach that builds on a
temporally smoothed $l^1$-regularized logistic regression formalism
that can be cast as standard convex-optimization problem and solved
efficiently using generic solvers scalable to large networks.

\item Working with scientific citation networks, hyperlinked web pages
and geographically tagged news articles, Chang and Blei develop a
\textit{Hierarchical Relational Model of Document Networks}. They develop
a hierarchical
model of both network structure where the attributes of each document
are its
words, and for each pair of documents, the model is their link as a
binary random variable that is conditioned on their contents. They
derive efficient inference and estimation algorithms based on
variational methods that take
advantage of sparsity and scale with the number of links.

\item Jank and Yahav focus on a dataset involving
30,000 auctions from one of the main consumer-to-consumer online
auction houses. They propose a novel measure of e-loyalty via the
associated network of transactions between
bidders and sellers. In \textit{E-Loyalty Networks in Online Auctions},
they employ ideas from functional principal component analysis to
derive, from this network, the distribution of
perceived loyalty of every individual seller and associated loyalty
scores. In the process, they confront the clustering feature of loyalty
networks, with a few high-volume sellers accounting for most of the
individual transactions.
\end{itemize}
Part II of this special section will explore another diverse collection
of network models and applications.
\iffalse

\fi
%

\printaddresses

\end{document}